\documentclass[prb,twocolumn,superscriptaddress,floatfix]{revtex4}

\usepackage{graphicx}

\begin{document}

\title{Angular dependence of the magnetic susceptibility in the
itinerant metamagnet Sr$_3$Ru$_2$O$_7$}

\author{S. A. Grigera}
\affiliation{School of Physics and Astronomy, University of
St. Andrews, St. Andrews KY16 9SS, UK.}  
\affiliation{School of Physics and Astronomy, University of
Birmingham, Birmingham B15 2TT, UK.}

\author{R. A. Borzi}
\affiliation{School of Physics and Astronomy, University of
St. Andrews, St. Andrews KY16 9SS, UK.}

\author{S. R. Julian} 
\affiliation{Cavendish Laboratory, Madingley Road, Cambridge CB3 OHE,
UK.}

\author{R. S. Perry}
\affiliation{Department of Physics, Kyoto University, Kyoto 606-8502,
Japan.}

\author{Y. Maeno} 
\affiliation{Department of Physics, Kyoto University, Kyoto 606-8502,
Japan.}
\affiliation{International Innovation Centre, Kyoto University, Kyoto
606-8501, Japan.}

\author{A. P. Mackenzie}
\affiliation{School of Physics and Astronomy, University of
St. Andrews, St. Andrews KY16 9SS, UK.}


\begin{abstract}
\phantom{a}
\vspace{4ex}
We report the results of a study of the differential magnetic
susceptibility of Sr$_3$Ru$_2$O$_7$ as a function of temperature and
magnetic fields applied at a series of angles to the $ab$ plane.  By
analysing the real and imaginary parts of the susceptibility, we
conclude that the field angle acts as a continuous tuning parameter
for the critical end-point to a line of first order metamagnetic phase
transitions.  The end-point sits at approximately 1.25K for fields
applied in the $ab$ plane, and is depressed to below 50 mK when the
field is aligned within 10$^\circ$ of the $c$ axis.
\end{abstract}

\date{\today}

\maketitle

\section{Introduction}

Understanding unusual behaviour of itinerant carriers in correlated
electron systems continues to be a major challenge of modern condensed
matter physics.  One of the most promising approaches is based on the
concept of quantum criticality, in which many of the metallic
properties are dominated by fluctuations associated with the presence
of a $T \to 0$ critical point
\cite{Her_76,Mil_93,Sac_99,Mat_98,Ste_01}.  Usually, quantum critical
points result from the tuning of second order phase transitions.
Recently, however, we have discussed the possibility that signatures
of quantum criticality observed in the bilayer ruthenate
Sr$_3$Ru$_2$O$_7$ are associated with a $T \to 0$ critical end-point
of a line of first-order metamagnetic phase transitions
\cite{Per_01,Gri_01,Gri_02,Mil_02}. The data on which this analysis
was based were taken with the field aligned with the crystallographic
$c$-axis, and it was argued that in this circumstance the critical
end-point existed naturally at or very near $T = 0$. However, no
explicit evidence had been obtained for a key feature of the
underlying picture, namely continuous tuning of a line of end-points
terminating a surface of first-order transitions.  In this paper we
report a study of the angular dependence of the differential magnetic
susceptibility, in which we demonstrate the existence of such a
surface.

The most general definition of metamagnetism is a superlinear rise in
magnetisation over a narrow range of applied magnetic field.  This can
occur in several circumstances.  The term was first used in the
context of antiferromagnetic insulators, for which the change in
magnetisation is due to field-induced `spin-flip' or `spin-flop'
transitions \cite{Str_77}.  A form of metamagnetism more relevant to
itinerant systems was first described by Wohlfarth and Rhodes
\cite{Woh_62}, who pointed out that strongly enhanced paramagnets on
the verge of ferromagnetism might undergo a phase transition involving
exchange splitting of the Fermi surface at some finite applied
magnetic field.  A characteristic of such systems is a strongly
enhanced paramagnetic susceptibility with a pronounced maximum as a
function of temperature, indicating the presence of a peak in the
density of states very near the Fermi level. In a mean field
description the application of a field leads to the Stoner criterion
for exchange splitting being satisfied, resulting in a metamagnetic
phase transition.

An itinerant electron metamagnetic phase transition of the kind
described above is usually first order, because the presence of the
magnetic field prevents a symmetry change -- the Fermi surface becomes
more polarised rather than spontaneously polarised \cite{Sac_02}.
These first order transitions have been seen in a number of systems,
as described in a recent review by Goto, Fukamichi and Yamada \cite
{Got_01}. In some cases ({\em e.g.} Co(S$_{1-x}$Se$_x$)$_2$, for $x <
0.1$ and La(Fe$_{1-x}$Si$_x$)$_{13}$, for $0.11 < x < 0.14$), the
systems undergo first order ferromagnetic phase transitions below some
temperature $T_c$.  As the temperature is raised higher than $T_c$ ,
the ferromagnetic transition at zero field is changed to a first order
metamagnetic transition.  In others ({\em e.g.}
Co(S$_{1-x}$Se$_x$)$_2$, for $x \geq$ 0.12), no ferromagnetism is
seen, and first-order metamagnetic transitions are observed as $T \to
0$.  In common with other first-order phase transitions, metamagnetic
transitions change from being first order to being crossovers at
sufficiently high temperature, with a critical end-point separating
the two regimes.  This end-point occurs at fairly high temperatures in
some of the best-studied systems (e.g.  $\approx$ 80 K in
Co(S$_{1-x}$Se$_x$)$_2$, for $x$ = 0.14, and $\approx$ 15 K in UCoAl).
It can be forced down in temperature by the application of hydrostatic
pressure (to below 10K in UCoAl at 1.2GPa)\cite{Mus_99}, but to our
knowledge, it has not been tuned to the $T \to 0$ in any metamagnet
studied so far under pressure.  However, any material has a natural
effective pressure related to its electronic structure.  In some
materials metamagnetic crossovers rather than phase transitions are
seen, even at low temperatures \cite{Flo_02} suggesting that chemical
pressure is capable of depressing the end-point to `negative
temperature'.

The above discussion raises the possibility that in some
circumstances, the chemical pressure of a material may be suitable to
tune a metamagnetic end-point to, or very close to, $T = 0$.  On the
basis of observations of transport properties, we have argued that
Sr$_3$Ru$_2$O$_7$ with the field aligned along the $c$ axis is an
example of such a situation.  If correct, this analysis suggests that
the material displays an unusual form of quantum criticality without a
symmetry-broken phase \cite{Gri_01, Gri_02}.  This would be an
important development in the on-going quest to understand quantum
critical points, so it is important to investigate the metamagnetism
of Sr$_3$Ru$_2$O$_7$ in more depth.  Of particular interest would be
the identification of a tuning parameter that would allow the
identification of first-order transitions terminating in a
finite-temperature critical point that could then be tuned smoothly to
$T = 0$.  In the work reported to date, the criticality has been
ascribed to such a non-symmetry breaking critical point, but because
it already sits at $T = 0$, no associated first order transition has
been explicitly seen.

Experience on other itinerant metamagnets shows that pressure
increases the characteristic field of the metamagnetic transition
while at the same time depressing the end-point temperature $T^*$ (see
ref \onlinecite{Got_01}).  For the field aligned parallel to the $c$
axis of Sr$_3$Ru$_2$O$_7$, pressure is therefore likely to favour a
crossover rather than a first-order transition with a finite
temperature end-point.  However, the natural anisotropy of the
metamagnetism suggests another route to achieving the desired tuning.
The metamagnetic fields are slightly different for the two field
orientations (5 T for $H \parallel ab$ and 7.8 T for $H \parallel c$)
\cite{Per_01}, but this is not the only feature of the anisotropy.
The low temperature magnetoresistance is qualitatively different for
the two field orientations, giving rise to the suspicion that for $H
\parallel ab$, the transition may be first order at $T = 0$.  In a
recent paper, Chiao {\em et al.} \cite{Chi_unp} have argued that
measurements of the differential magnetic susceptibility for $H
\parallel ab$ support the hypothesis of a first-order transition.

Here, we report the results of a detailed study of the differential
magnetic susceptibility as a function of temperature, magnetic field
and magnetic field angle.  We combine observations of the real and
imaginary parts of the susceptibility and of hysteresis to present
strong evidence that the phase diagram contains a surface of first-order
phase transitions in ($H$, $T$, $\theta$) space.  The line of critical
end-points that terminates this surface is tuned continuously towards
zero temperature as the field is rotated away from the plane, falling
below 50 mK (the lowest temperature studied) when the angle is within
10$^\circ$ of the $c$ axis.  We further demonstrate that even the
relatively low frequency ( $\approx 80$ Hz) employed for these
a.c. measurements is not a good approximation to the static limit.
Care must therefore be taken before drawing conclusions on the nature
of the critical point from the field and temperature dependence of the
susceptibility in its vicinity.

\section{Experimental}

The experiments described here were performed in dilution
refrigerators in Cambridge and St. Andrews.  The sample studied was a
small piece of approximate dimension (1 x 1.1 x 0.6) mm$^3$ cut from a
single crystal of Sr$_3$Ru$_2$O$_7$ grown in Kyoto whose residual
resistivity had been measured to be 2.4 $\mu\Omega$cm. It was
thermally grounded to the mixing chamber by electrical coupling
through gold and copper wires. In order to work at temperatures above
1.1 K the normal regime of operation of the dilution fridge was
changed by varying the $^3$He/$^4$He mixture composition. The
differential susceptibility was measured using a drive field of 3.3 x
10$^{-5}$ T r.m.s., and counterwound pickup coils each consisting of
approximately 1000 turns of 12 $\mu$m diameter copper wire.  For most
of the experiments, an excitation frequency of 79.96 Hz was employed.
Low temperature transformers mounted on the 1K pot of the dilution
refrigerators were used throughout to provide an initial signal boost
of approximately a factor of 100.  When the frequency dependence was
studied during the later stages of the project, the frequency
dependence of the transformer was explicitly calibrated rather than
relying on generic performance data.  This ensures that we have
confidence in the relative magnitude of the signals presented here.

Direct measurement of the empty-coil signal as a function of angle
established that at all angles studied, the background varied smoothly
with field through the field region of interest (between 4.5 and 8.5
tesla).  The data shown in Figs. 1-3 were obtained by subtracting a
second order polynomial which modelled the combination of the
empty coil background and the weakly varying non-metamagnetic
susceptibility from the raw signal.  The data presented are therefore
the extra susceptibility due to the metamagnetism of
Sr$_3$Ru$_2$O$_7$.

The absolute magnitude was more difficult to measure with high
accuracy.  It was calibrated by measuring the signal with and without
the sample at fields between 1 and 3 T applied parallel to the $ab$
plane.  The direct background subtraction gave the expected
field-independent susceptibility in this field range, and this was
calibrated against the known value \cite{Per_01,Ike_00}.  The dilution
refrigerator had to be removed from and then replaced in the cryostat
in order to perform the empty run.  The accuracy of our absolute
calibration depends on the stability of the background signal
throughout this process, and on the assumption that a measurement at
approximately 80 Hz is equivalent to one at quasi-d.c. in this field
range.  However, we stress again that these concerns do not apply to
the relative signal magnitudes that we report.

\section{Results} 

A representative sample of the large data set acquired during the
course of this work is shown in Fig. 1.  The real part of the dynamic
susceptibility is plotted as a function of a swept d.c. field applied
at an angle of 40$^\circ$ to the $ab$ plane.  At this field angle, the
data contain a main peak at approximately 5.7 T, and a much weaker
second peak at approximately 6.7 T.  The second peak washes out as the
temperature is increased from 100 mK.  This and a further minor peak
exist when the field is applied parallel to the $ab$-plane, but both
disappear rapidly with temperature and increasing field angle.  Here,
we concentrate our attention on the major peak.  This peak {\em grows}
as the temperature is increased from 100 mK, reaching a pronounced
maximum at approximately 1 K before decreasing in magnitude at higher
temperatures.  The characteristic field of the maximum is also weakly
temperature-dependent, decreasing by approximately 0.05 tesla as the
temperature is raised from 100 mK to 1.4 K.
%
\begin{figure}
\includegraphics[width=\columnwidth]{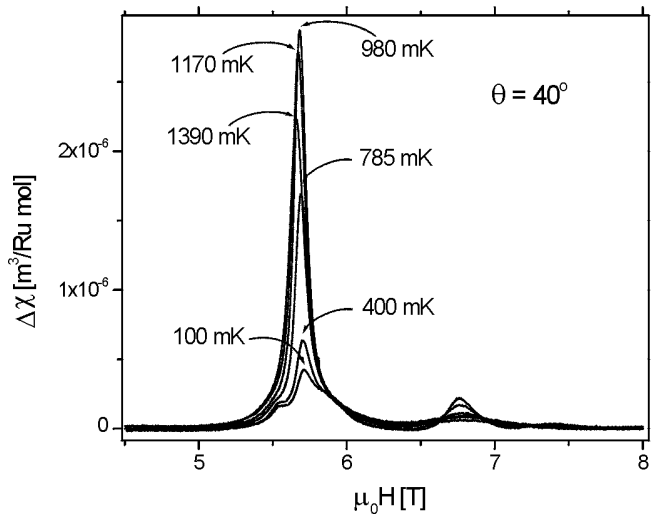}
\caption[]{The real part of the differential magnetic susceptibility
through the metamagnetic transition in Sr$_3$Ru$_2$O$_7$.  A small
oscillating field (3.3 x 10$^{-5}$ T r.m.s., 79.96 Hz) is combined
with a swept d.c. field, applied in this case at 40$^\circ$ to the
$ab$ plane.  The main peak shows a pronounced maximum at approximately
1 K, falling away in magnitude for temperatures higher or lower than
this.  All data were taken in an increasing d.c. field. The data are
quoted as $\Delta\chi$ because of the background subtraction procedure
described in the text.  }
\end{figure}
%
%
%
\begin{figure}[!]
\includegraphics[width=\columnwidth]{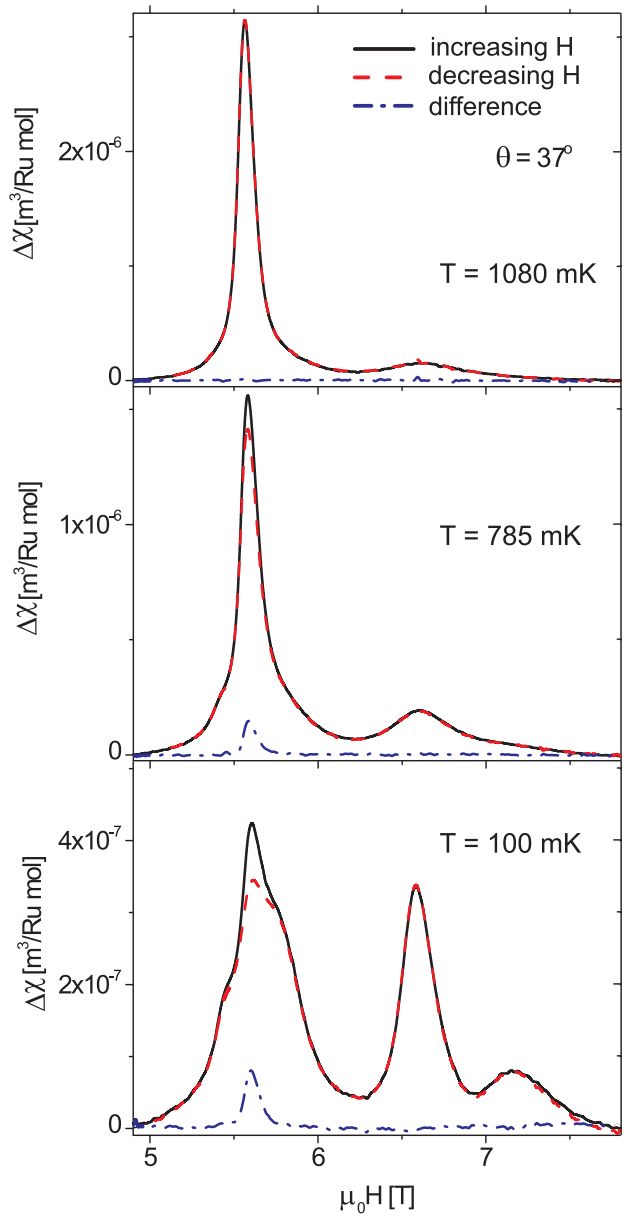}
\caption{(colour online) The real part of the differential magnetic
susceptibility through the metamagnetic transition in
Sr$_3$Ru$_2$O$_7$.  At 785 mK and 100 mK, hysteresis is seen in the
most prominent peak as the d.c. field, applied at approximately
40$^\circ$ to the $ab$ plane, is increased (black solid line) and
decreased (red dashed line).  At 1080 mK the hysteresis has completely
disappeared.  In each panel, the blue dot-dashed line gives the
difference between the results for the increasing and decreasing
field.  In this case the frequency of the oscillating field was 63.54
Hz. The data are quoted as $\Delta\chi$ because of the background
subtraction procedure described in the text.  }
\end{figure}
%
%
%
\begin{figure*}[!]
\includegraphics[width=17cm]{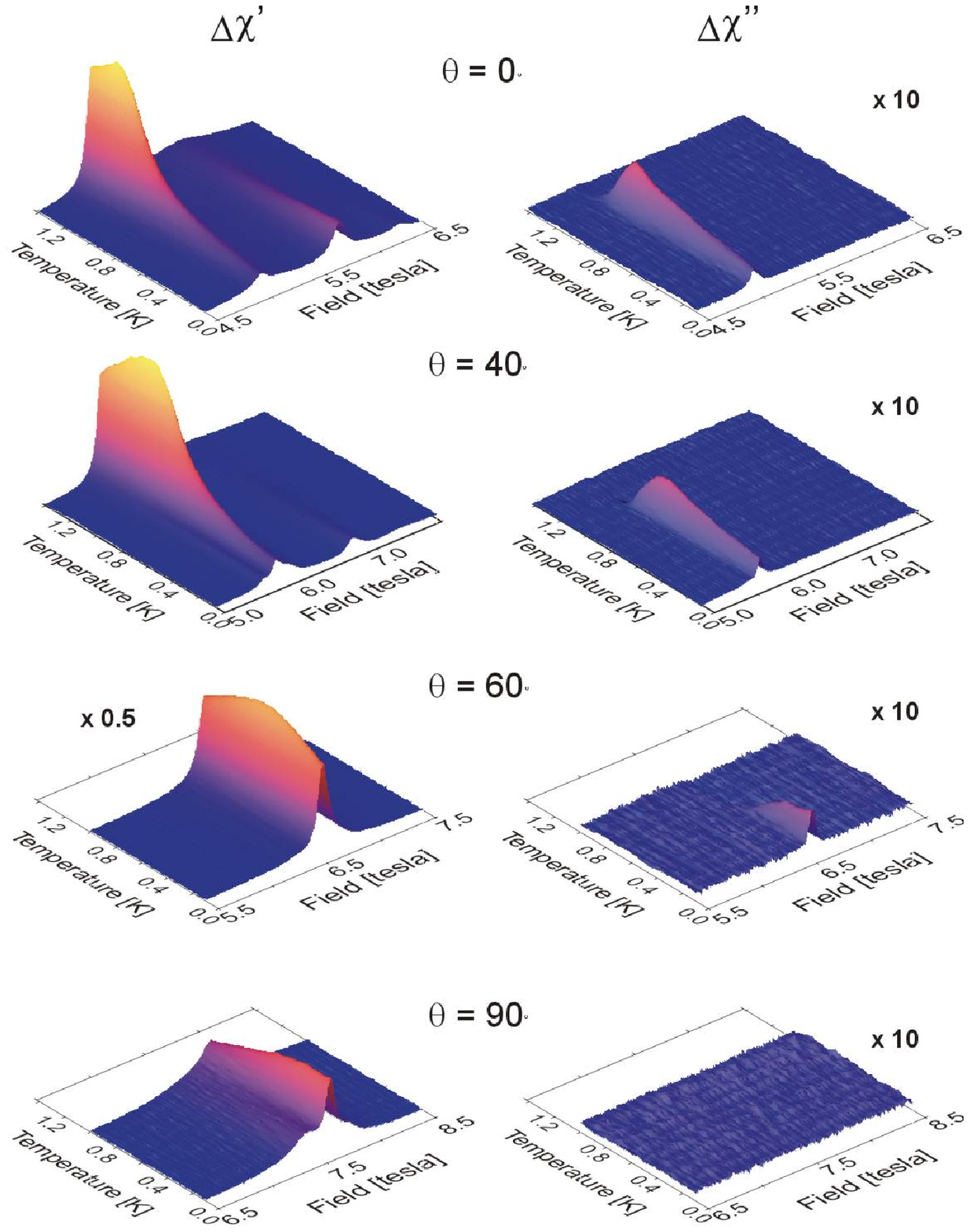}
\caption{(colour online) The real and imaginary parts ($\Delta\chi'$
and $\Delta\chi''$ respectively) of the differential susceptibility of
Sr$_3$Ru$_2$O$_7$ shown in a three-dimensional format with amplitude
plotted against temperature and magnetic field, at a series of field
angles.  In all cases the frequency of measurement was 79.96 Hz, and
the data were taken in an increasing d.c.  field. An excellent
correlation is seen between the overall maximum in the real part of
the susceptibility and the appearance at low temperatures of a peak in
the imaginary part, indicative of the onset of dissipation.  For the
field applied parallel to the $c$ axis, the maximum is at the lowest
temperature reached, and no peak at all is seen in the imaginary
component. The imaginary parts of the susceptibility are multiplied by
a factor of 10, and at $\theta = 60^\circ$ the real part by a factor
0.5.}
\end{figure*}

The behaviour of the real part of $\chi$ shown in Fig. 1 is suggestive
of a line of first order transitions ending in a critical point at a
temperature $T^*$ at which the susceptibility is maximised.  The
argument is as follows.  At high temperatures, above that of the
critical point, the susceptibility behaves like that of a crossover.
As the temperature is decreased towards $T^*$, the peak susceptibility
increases before being cut off by finite frequency or finite size
effects.  Below $T^*$, the dynamical response becomes sensitive to the
physics of a first-order magnetic transition such as domain wall
movement, causing the susceptibility to decrease.  Previous work on
systems such as Co[(CH$_3$)$_3$NH]Cl$_3~\cdot$2H$_2$O] gives a precedent
for this kind of behaviour, and also indicates that non-zero frequency
effects can play an important role even for low excitation frequencies
of the kind ($\approx 80$ Hz) employed here \cite{Duy_82}.

The above argument suggests that the data in Fig. 1 are consistent
with the existence of a line of first-order transitions terminating in
a critical point (for this field angle) at approximately 1 K, but they
certainly cannot be regarded as proof of that hypothesis.  Firmer
evidence comes from a study of hysteresis, shown in Fig. 2.  If the
a.c. response of the system is studied while the main d.c. field is
swept in opposite directions, a clear hysteretic response in the
magnitude of $\Delta\chi$, but not in the value of $H$ at the peak, is
seen for temperatures less than that at which the susceptibility is
maximum\cite{com}.  The existence of hysteresis for temperatures below
$T^*$ is the first strong support provided by our data to the idea of
a first-order metamagnetic transition in this range of temperatures.

A second and related piece of evidence comes from consideration of a
dissipative component in the a.c. signal itself.  A wider sample of
our total data set is shown in Fig.  3.  In order to present the
results of many field sweeps at each angle in a single plot, we adopt
a colour-coded three dimensional format in which the magnitude of the
dynamical susceptibility is plotted for an area of the ($H$, $T$)
plane.  Data are shown for field angles $\theta$ = 0$^\circ$,
40$^\circ$, 60$^\circ$ and 90$^\circ$ from the $ab$-plane.  In each
case, the plots were constructed using field sweeps at discrete
temperatures at approximately 100 mK intervals from 100 mK to the
highest temperature shown.  The second graph on the left of Fig. 3
therefore includes the data shown in Fig. 1 combined with the results
of eight other field sweeps.  The data for the real part of $\chi$
show a maximum susceptibility ($T^*$, $H^*$) for which $T^*$ drops and
$H^*$ rises as $\theta$ increases.  For $\theta \agt 80^\circ$, the
highest observed value of $T^*$ is seen at the lowest temperature
studied.  The most significant aspect of this form of data
presentation is that it gives a good visualisation of the correlation
between the behaviour of the real part of the dynamical susceptibility
and that of the imaginary part.  Hysteresis of the kind shown in
Fig. 2 would be expected to be accompanied by dissipation, and
dissipation leads to the appearance of a feature in the imaginary part
of an a.c. response.  As can be seen from Fig. 3, we observe a direct
correlation between the characteristic temperature of the maximum in
the real part of the dynamical susceptibility and that below which a
dissipation peak appears in the imaginary part.

It is hard to explain data of the kind shown in Fig. 3 without
postulating the existence of a first-order metamagnetic transition
below the empirically-defined characteristic temperature $T^*$.
Before discussing the meaning of the characteristic temperature and
field ($T^*$, $H^*$) further, we present some more empirical findings
relating to their angle dependence. The data shown in Fig. 3 are again
only a sample from the complete data set acquired during our
experiment, in which data were taken at a series of angles between
0$^\circ$ and 90$^\circ$ and at temperatures from 1400 mK to 50 mK.
A three-dimensional representation of the experimental phase diagram
is given in Fig. 4.  In this figure, the shaded surface is defined as
the locus of points at which a peak is observed in the imaginary part
of the dynamic susceptibility.  Since this peak is linked to the
dissipative response of the system, our interpretation of the shaded
area is a surface of first order metamagnetic phase transitions in the
three-dimensional ($H$, $T$, $\theta$) space.  The solid line was
obtained from the series of points at which the real part of the
susceptibility has its overall maximum (clearly visible in each of the
panels on the left-hand side of Fig. 3), which we will argue below to
be a line of metamagnetic critical points.
%
\begin{figure}
\includegraphics[width=\columnwidth]{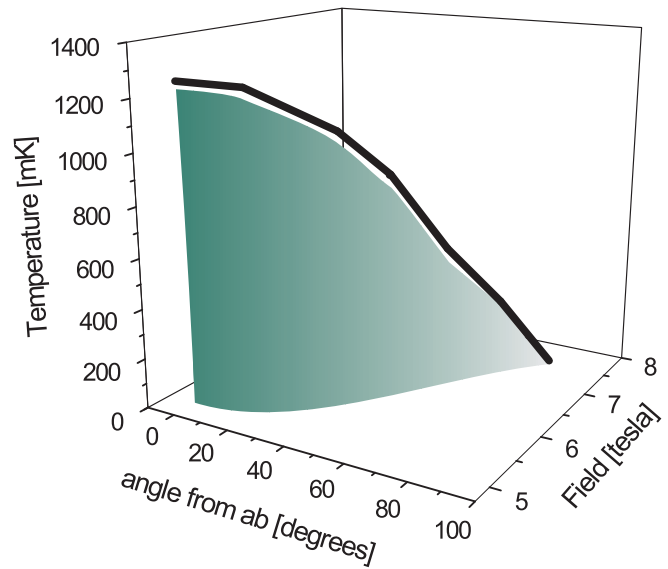}
\caption[]{(colour online) The phase diagram inferred from the whole
body of susceptibility measurements reported in this paper.  The
shaded region represents the locus of points at which the imaginary
part of the susceptibility has a peak, while the solid line is the
locus of the overall maxima seen in the real part.  Increasing
magnetic field sweeps through the metamagnetic transition were taken
at least every 100 mK, at a series of field angles as defined by the
discrete points in Fig. 5.  }
\end{figure}

Fig. 4 is useful as an overall summary of our main experimental
findings, but it can be difficult to read off some of the detailed
information that has been obtained.  For this reason, we present two
projections from the figure in the main parts of Figs. 5 and 6.  In
Fig. 5 we show the projection of the `line of critical points' onto
the ($T$, $\theta$) plane, showing how the characteristic temperature
$T^*$ is depressed as the field angle is changed. In the main part of
Fig. 6 the line is projected onto the ($H$, $\theta$) plane, giving a
representation of the angle dependence of the critical metamagnetic
field.  The inset to Fig. 5 shows that although the shaded surface in
Fig. 4 is almost vertical, the slight curvature that exists is
`convex'.  Here the points show the ($T^*$, $H^*$) values of the
overall maximum of the real part of $\chi$ at each of the two
angles. The lines are two cuts from the plane at constant $\theta$
values of 50$^\circ$ and 60$^\circ$, so the inset confirms that, like
the peak in the real part of $\chi$ ({\em e.g.} Fig. 1), the peak in
the imaginary part falls to slightly lower field as temperature
increases.  The inset to Fig. 6 will be described in the next section.
%
\begin{figure}
\includegraphics[width=\columnwidth]{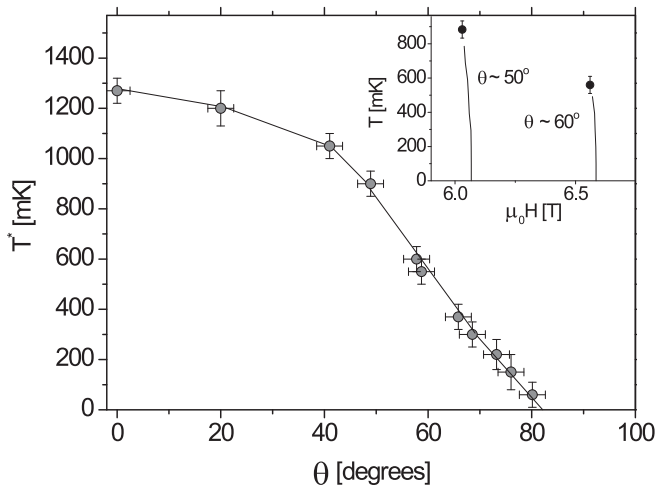}
\caption[]{ The critical temperature $T^*$ (defined in the text) as a
function of angle.  The data are a discrete projection of the solid
line from Fig. 4 onto the temperature -- angle plane.  For angles
above $\approx 80^\circ$ the susceptibility is a monotonically increasing
function to the lowest temperatures measured (50mK). The inset shows
the field and temperature dependence at angles of 50$^\circ$ and
60$^\circ$ of the peak in the imaginary part of $\chi$, which we
interpret as defining lines of first order transitions (solid lines).
The solid points are the position in temperature and field of the
overall maximum of the real part of $\chi$ at these angles.}
\end{figure}

\section{Discussion}

The principal findings of our work, summarised in Figs. 1 -- 6, are
two-fold.  First, we believe that our results give very good evidence
for the existence of a surface of first-order metamagnetic phase
transitions in Sr$_3$Ru$_2$O$_7$.  Varying the combination of field
and field angle terminates this surface at different temperatures $T^*$,
with $T^*$ falling from approximately 1.25 K for $H \parallel ab$ to
less than our base temperature of 50 mK for $H \parallel c$.  Second,
the fact that field angle plays the role of a continuous tuning
parameter for this system is a significant finding, which should open
the way to future experiments involving fine adjustment of the balance
between thermal and quantum fluctuations.  One example of the effects
of such a balance seems already to be observable in the data. The fall
in the first-order metamagnetic transition field with increasing
temperature (inset to Fig. 5) is in contrast with observations on
other ferromagnetic itinerant metamagnets with higher critical
temperatures \cite{Got_01} and with the predictions of a classical
model\cite{Yam_93}; the disordering effect of thermal fluctuations
would imply that {\em higher} fields are needed to reach the state of
enhanced magnetisation.  This apparent paradox is no longer present
when both thermal and quantum fluctuations are taken into account
\cite{Mil_02}.  At very low temperatures the latter dominate and the
tendency is reversed.

The basic conclusions reached here about the behaviour of $T^*$ are
also consistent with previous work on this material.  For $H \parallel
c$, transport measurements show a steep increase of the $A$
coefficient which measures the strength of the quadratic term in the
resistivity.  This suggests a diverging quasiparticle mass as the
metamagnetic field is approached at low temperatures, as would be the
case if there were a very low or zero temperature (quantum)
metamagnetic critical point \cite{Gri_01}.  

A puzzle from the transport study was the appearance of anomalous
powers higher than two in the $T$-dependent resistivity in a very
narrow range of fields near the metamagnetic field \cite{Gri_01}.  In
the absence of other data, one possibility for this anomalous
behaviour was that the system was entering a region of first-order
physics in this field range \cite{Gri_01,Aep_01}.  The data presented
in Fig. 4 give no support to this idea.  On the contrary, the unusual
transport properties are seen in precisely a region of the phase
diagram where we see no susceptibility evidence for first-order
transitions.
%
\begin{figure}
\includegraphics[width=\columnwidth]{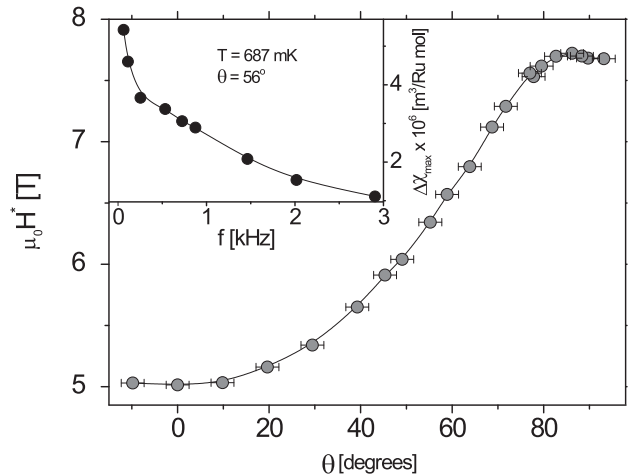}
\caption[]{ The critical field $H^*$ (defined in the text) as a
function of angle. The data are a discrete projection of the solid
line from Fig. 4 onto the field -- angle plane.  The inset shows the
frequency dependence of the peak in the real part of $\chi$ for $T^*$
at a field angle of 56$^\circ$.  }
\end{figure}

The data discussed so far give convincing evidence that a temperature,
field and field- angle-dependent critical point must exist in
Sr$_3$Ru$_2$O$_7$, because the surface of first-order transitions that
has been identified must end in a line of critical points.  Open
questions remain, however, about the precise nature of these critical
points.  A central issue concerning metamagnetic criticality is
whether a simple picture based around long-wavelength `ferromagnetic'
fluctuations \cite{Mil_02,Yam_93} gives an adequate description of the
physics.  For instance, it has been argued that even though a surface of
first-order transitions can be observed by a long-wavelength probe
like the magnetic susceptibility employed here, it is possible that
its end-point might involve criticality of fluctuations at a much
shorter wavelength \cite{Chi_unp}.  The metamagnetic crossover in
CeRu$_2$Si$_2$ has been argued to originate from such
`antiferromagnetic-like' fluctuations \cite{Flo_02}, and
nesting-related short-wavelength fluctuations have been observed by
inelastic neutron scattering in Sr$_3$Ru$_2$O$_7$ \cite{Cap_unp}.

At first sight, our susceptibility data do not appear to favour the
simple long-wavelength picture \cite{Mil_02}, because although an
overall maximum is observed in the real part of $\chi$ at $T^*$, the
divergence that such a model would predict is not seen.  In real
systems, divergences near phase transitions are cut off by finite size
effects, disorder and (near magnetic transitions) demagnetisation
\cite{Duy_82}.  We have made estimates of the magnitude of each of
these effects, and have concluded that none of them would change a
putative divergence to the much weaker feature shown in Fig. 3.
However, we have discovered that although low, the finite frequency of
these differential susceptibility measurements has a much stronger
effect than we had anticipated.  In the inset to Fig. 6, we show the
frequency dependence of the peak in the real part of $\chi$ at $T^*$
(approximately 700 mK) for a field angle of approximately 56$^\circ$ .
The peak height is seen to be rising steeply through the frequency
range including that ($\approx 80$ Hz) of most of our measurements.
Strong frequency dependences are also seen below $T^*$, although with
a different functional form (data not shown).  Although the origin is
magnetic in both cases, there are differences.  Below $T^*$ domain
wall motion plays a dominant role, whereas at and above $T^*$,
critical fluctuations are the main source.  Although the sample
studied here is very pure in comparison to most metallic oxides, the
disorder that does exist may be playing a role in this unexpectedly
strong frequency dependence\cite{Ros_99}.  The critical slowing down
expected in the vicinity of a critical point can be greatly enhanced
by the presence of disorder, with even very small quantities of it
acting as `pinning centres' for fluctuations.  This pinning is in turn
enhanced by the (low) scale of temperature at which the transitions
take place.

The strength of the frequency dependence shown in the inset to Fig. 6
means that detailed analyses based on the field or temperature
dependence of the {\em amplitude} of the data shown in Figs. 1 -- 3
must be done with great care.  The data that we have presented are a
fixed frequency snap-shot of a more complicated overall situation, and
that snap-shot is not a good approximation to the zero frequency
limit.  We therefore believe that no firm conclusions can be drawn
based on the weakness of the overall maximum seen in the real part of
$\chi$.  In particular, one certainly cannot rule out a genuine
divergence in the long wavelength limit on the basis of a study such
as this.

To our knowledge, previous studies of metamagnetism using finite
frequency a.c.  methods have been relatively rare.  Most reports of
susceptibility have been obtained from the results of measurements of
quasi-d.c. magnetisation (in these studies the sweep rate of the
d.c. field determines the characteristic time
constant)\cite{Str_77,Got_01}.  The work reported here emphasises that
a.c. methods are a very sensitive means of detecting first-order
behaviour and its temperature dependence \cite{Duy_82}. They provide,
in addition to the simplicity of their implementation, a way to study
dynamical properties of the system.  It is clear that a more thorough
study of frequency dependence would be desirable.  Preliminary
measurements have shown that the determination of $T^*$ itself has
some frequency dependence, which must be taken into account if
elevated frequencies in the kHz range are employed.  Below 100 Hz,
however, the effect is weak, so that we believe that the phase
diagram presented in Fig. 4 is a good approximation to the results
that would be obtained in the zero frequency limit.

Inspection of Fig. 4 shows that the inferred line of critical points
passes below 50 mK at an angle of approximately $80^\circ$, with a
gradient that suggests that it would reach $T = 0$ at approximately
$85^\circ$ rather than $90^\circ$.  There is no a priori reason to
favour one angle over the other, or to assume that the critical point
sits at $T = 0$ only at one angle and not at a range of angles.  On
the basis of the data presented here, we cannot draw a firm conclusion
either way.  We also note that it may never be possible to do so.  If
the anomalous transport behaviour reported very near $H^*$ for $H
\parallel c$ is due to the system entering some new state in the
vicinity of the quantum critical point, the zero temperature critical
point itself may not prove to be experimentally accessible.  We plan
more work in this area.

All of the discussion so far has concentrated on the most prominent
peak seen in Figs. 1--3.  No first-order behaviour is associated with
either of the smaller peaks seen at higher fields for low $\theta$.
On the basis of the data presented here, there is no reason to state
that either of this peaks correspond to magnetic quantum critical
points.  Rather, the rapid disappearance with $\theta$ makes
interpretation in terms of crossovers more likely.  However, a
detailed study of the frequency dependence will be necessary to settle
this issue one way or the other.

Finally, we briefly raise the issue of the {\em origin} of the angular
dependence.  So far, we have concentrated on establishing the
empirical fact that field angle acts as a continuous tuning parameter
for the metamagnetism of Sr$_3$Ru$_2$O$_7$ rather than discussing why
that should be the case.  Our results show that the metamagnetic
transition presents both quantitative and qualitative changes as a
function of the field angle that are not compatible with a simple
projection of the magnetic field into the principal axes.  It is known
that lattice distortions have a pronounced effect on the magnetic
properties of ruthenates \cite{Sha_00,Fri_01}, and that in particular
the metamagnetic transition in Sr$_3$Ru$_2$O$_7$ is strongly affected
by the application of hydrostatic and uniaxial pressure
\cite{Chi_02,Ike_unp}.  All this points towards magnetostriction
effects being an important factor in the mechanisms of the
metamagnetic transition in Sr$_3$Ru$_2$O$_7$.  Such effects are known
to play an important role in the metamagnetism of other materials
\cite{Sou_Ce}.  If there is a non-trivial angle-dependent
magnetostriction associated with the metamagnetic transition in
Sr$_3$Ru$_2$O$_7$, there would be no reason for the effects of a
changing field angle to be felt only as a projection.  A puzzling
feature of the magnetism of this material has been the fact that
although there is some anisotropy of the metamagnetism, the low-
temperature weak-field magnetic susceptibility is remarkably isotropic
\cite{Ike_00}.  A scenario in which the anisotropy of the
metamagnetism is related to magnetostriction promoted by higher
applied fields might give a natural explanation for these apparently
contradictory observations.  A further appealing aspect of this
picture is a natural analogy between changing field angle and changing
pressure.  In this sense the empirically determined phase diagram
shown in Fig. 4 would have a close relation to the surface of
pressure-induced metamagnetic phase transitions that results from
Ginzburg-Landau treatments of varying sophistication
\cite{Gri_02,Got_01,Pfl_01}.

\section{Conclusions}

In summary, we have used measurements of the differential dynamic
susceptibility to demonstrate that changing magnetic field angle
represents a continuous tuning parameter for the metamagnetism of
Sr$_3$Ru$_2$O$_7$.  The empirical phase diagram contains a surface of
first-order metamagnetic transitions whose end-point temperature
$T^*$ can be tuned from 1.25 K for fields parallel to the $ab$ plane
to below 50 mK for fields applied parallel to the $c$ axis.  These
and previous transport studies are consistent with the existence of a
quantum critical end-point for the latter field orientation.  The work
shows that a.c. susceptibility can be a sensitive and powerful probe
of the classical- quantum crossover in metamagnetism, but that a
fuller study of the frequency dependence is desirable.

\begin{acknowledgements}

We are grateful to L.M. Galvin for contributions to the early stages
of this project, and to P. Coleman, E. M. Forgan, G.G. Lonzarich,
A.J. Millis, A. Rosch, P. Riedi, S. Sachdev and A.J. Schofield for
useful discussions.  We thank the Leverhulme Trust, the UK EPSRC and
the Royal Society for their support of the work.

\end{acknowledgements}


\begin{thebibliography}{99}
\bibitem{com} The hysteretic behaviour that we see with these
a.c. techniques differs from that usually observed in metamagnetic
transitions studied with d.c. techniques (See, for example ref
\onlinecite{Got_01}). The a.c. measurements include effects due to the
dynamics of the first order transition, which is markedly dependent on
whether the change is between a state of low and high magnetisation or
vice versa.
\end{thebibliography}
\end{document}